\documentclass[twocolumn,preprintnumbers,a4,prb]{revtex4}
\usepackage{graphicx}%
\usepackage{dcolumn}
\usepackage{amsmath}
\usepackage{amssymb}
\usepackage{bm}
\makeatletter
\def\btt#1{\texttt{\@backslashchar#1}}%
\DeclareRobustCommand\bblash{\btt{\@backslashchar}}%
\makeatother

\begin{document}

\preprint{AlMgB2-i.tex}

\title{A magnetization and $^{11}$B NMR study of Mg$_{1-x}$Al$_x$B$_2$ superconductors}
\author{M. Pissas$^1$,G. Papavassiliou$^1$,M. Karayanni$^1$,M. Fardis$^1$,
I. Maurin$^2$, I. Margiolaki$^2$, K. Prassides$^{1,2}$, C. Christides$^{1,3}$}
\affiliation{$^1$Institute of Materials Science, NCSR,  Demokritos, 153 10 Aghia Paraskevi, Athens, Greece\\
$^2$School of Chemistry, Physics and Environmental Science, 
University of Sussex, Brighton BN1 9QJ, UK\\
$^3$Department of Engineering Sciences, School of Engineering, University of Patras, 26110 Patras, Greece}

\date{\today}
\begin{abstract}
We demonstrate for the first time the magnetic field distribution of the pure vortex state in lightly 
doped Mg$_{1-x}$Al$_x$B$_2$ ($x\leq 0.025$) powder samples, by using $^{11}$B NMR in magnetic fields 
of $23.5$ and  $47$ kOe. The magnetic field distribution at $T=5$ K is Al-doping dependent, revealing a 
considerable decrease of anisotropy in respect to pure MgB$_2$. This result correlates nicely with magnetization 
measurements and is consistent with $\sigma$-band hole driven superconductivity for MgB$_2$.
\end{abstract}
\maketitle

The synthesis of MgB$_2$ had been reported in 1954\cite{jones54}
but only recently Nagamatsu {\it at al.} \cite{nagamatsu01} discovered that this compound is
a superconductor with a surprisingly high $T_c\approx 39$ K.
At first it was suggested\cite{kortus01} that a BCS-type mechanism with strong electron-phonon 
coupling and high phonon energy of the light boron atoms can be responsible for the observed 
high-$T_c$. This is based on the observation of the isotope effect\cite{budko01a} on $T_c$ 
and a strong negative pressure coefficient of $T_c$.\cite{pressure} 
Alternatively, Hirsch\cite{hirsch01} proposed a "universal" mechanism where superconductivity 
in MgB$_2$ is driven by the pairing of dressed holes. 
Electronic band structure calculations \cite{kortus01,suzuki01,An01,belashchenko01,satta01} 
indicate that in MgB$_2$ the charge carriers are situated in two bands derived from the 
$\sigma$-bonding $p_{x,y}$-orbitals of boron, which are essentially two-dimensional ($2D$),
and in one electron and one hole bands derived from the $\pi$-bonding $p_z$-orbitals of boron.
Both, $\sigma$ and $\pi$ bands have strong in-plane dispersion due to the large overlap between 
all $p$ orbitals of neighboring boron atoms. Despite some diversities in these models, there is 
a general agreement \cite{kortus01,hirsch01,An01,belashchenko01,Kong01,Singh01} that the 
key point  for superconductivity in MgB$_2$ is the $2D$ $\sigma$ band of $p_{x,y}$ orbitals within the 
boron layers, and the delocalized metallic-type bonding between these layers. These calculations 
predict \cite{An01,Lima01,Wang01,Bouqet01,kortus01} a strong anisotropy in the Fermi surface 
(and possibly in the electron phonon coupling) that is consistent with the observed 
\cite{Lima01,Patnaik,simon01} anisotropy in $H_{c2}$. Specifically, the anisotropic ratio: 
$\gamma$=$H_{c2}^{ab}$/$H_{c2}^c$, was found \cite{Lima01,simon01,budko01b} to be between $1.7$ and 
$6$, depending on the material and the experimental method.
\begin{figure}[tbp] \centering
\includegraphics[angle=0,width=6.0cm]{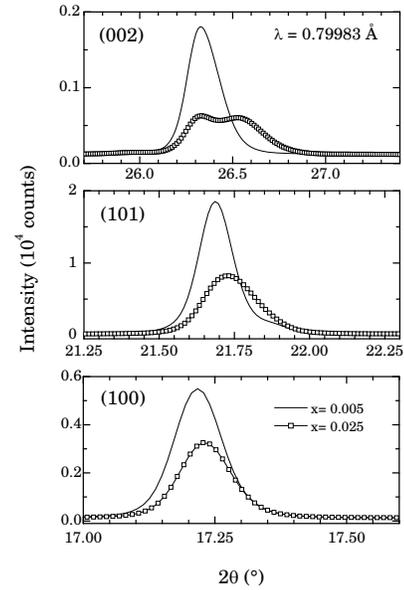}
\caption{Synchrotron X-ray diffraction ($\lambda$= 0.79983 \AA)
profiles showing the $(002)$, $(101)$ and $(100)$ reflections for the
Mg$_{1-x}$Al$_x$B$_2$ samples with $x=0.005$ and 0.025. }
\label{fig1}%
\end{figure}
\begin{figure}[hbp] \centering
\includegraphics[angle=0,width=8.0cm]{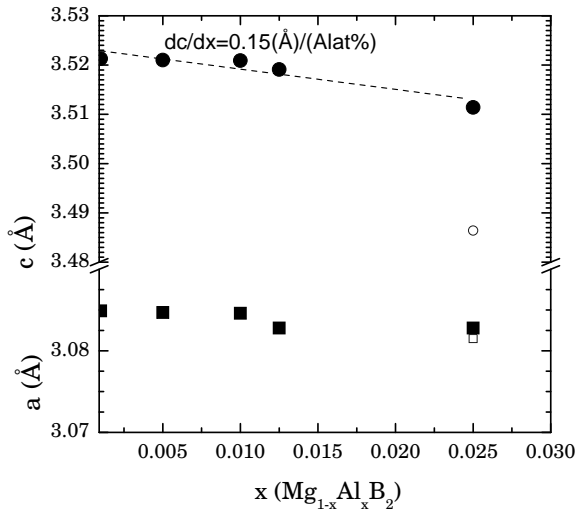}
\caption{Variation of the hexagonal unit cell parameters of  
Mg$_{1-x}$Al$_x$B$_2$ for $0\leq x \leq 0.025$.
Solid circles and squares correspond to the $c$- and $a$-axis of the
Mg$_{1-x}$Al$_x$B$_2$ phase, respectively. The corresponding open symbols for $x=0.025$
are the cell constants of the second (Mg$_{1-x}$Al$_x$B$_2$)$^/$ phase (see main text)
 }
\label{fig2}%
\end{figure}

In view of this description, measurements on electron- or hole-doped MgB$_2$ 
are of interest as they may help our understanding of how
the electronic density-of-states and the Fermi surface depend on doping. Al substitution for Mg in 
Mg$_{1-x}$Al$_x$B$_2$ provides a way for electron doping.\cite{Slusky01,xiang} The similarity 
of the calculated electronic density of states between MgB$_2$ and AlB$_2$ indicates that 
doping results in simple filling of the available electronic states, with one electron donated per Al.
\cite{suzuki01,satta01,belashchenko01} 
A very first study of Al doped MgB$_2$ showed\cite{Slusky01} that $T_c$ is slightly 
supressed for $x\leq 0.1$. However, band structure calculations show \cite{kortus01} that there 
is a sharp drop in the density of states of MgB$_2$ at only slightly higher electron 
concentrations. Suzuki {\it et al.} \cite{suzuki01,satta01} predict that in Mg$_{1-x}$Al$_x$B$_2$ the 
concentration of $\sigma$ holes varies with $x$ as $n_h=(0.8-1.4x)\times 10^{22}$ cm$^{-3}$, 
leading to $n_h=0$ for $x\approx 0.6$. For $0.1 \leq x \leq 0.25$, a two phase mixture is 
formed, whereas for $x>$0.25 a single non-superconducting phase is detected. The detrimental 
effect of doping on $T_c$ in Mg$_{1-x}$Al$_x$B$_2$ can be explained within the BCS model, as
it increases the Fermi energy ($E_F$) and decreases 
the density of states $N(E_F)$.
Besides, thermoelectric power and resistivity measurements show \cite{Lorenz01} that 
Mg$_{1-x}$Al$_x$B$_2$ alloys are hole-type normal metals. 
In order to analyze trends associated with the band filling and their relation to the loss of supeconductivity, 
we have performed a detailed study of Mg$_{1-x}$Al$_x$B$_2$ ($ 0\leq x \leq 0.1$)
using structural, magnetic and $^{11}$B NMR line shape measurements. 
Powder samples with nominal composition 
Mg$_{1-x}$Al$_x$B$_2$ ($0\leq x \leq 1$) were prepared by liquid-vapor to 
solid reaction as described elsewhere. \cite{Pissas01} 
Synchrotron X-ray powder diffraction measurements were performed on
Mg$_{1-x}$Al$_x$B$_2$ samples, sealed in
thin-wall glass capillaries $0.5$ mm in diameter at $295$ K. \cite{Comment1} 
Inspection of Figure \ref{fig1} which shows parts of the XRD patterns 
of the $x=0.005$ and $0.025$ samples shows that while 
Mg$_{1.995}$Al$_{0.005}$B$_2$ is single phase, a clear splitting of the
$(002)$ reflection is observed for Mg$_{1.975}$Al$_{0.025}$B$_2$. This implies 
the onset of macroscopic phase separation with increasing Al content. A similar
observation has been reported for C-doped MgB$_2$ compositions   
in which the carbon miscibility is also very small, $x<$ 0.04.\cite{MgBC} 
For this reason we restrict our NMR study
only to Mg$_{1-x}$Al$_{x}$B$_2$ samples with $x\leq 0.025$.
The deduced lattice parameters are plotted in Fig. \ref{fig2}. 
For $x< 0.025$, the $c$-axis exhibits a negative 
slope $dc/dx\approx(-0.2$ \AA/at \% Al), whereas the in-plane $a$-axis remains
nearly constant. 
For $x\geq 0.025$, the coexisting phases differ mainly in their interlayer
lattice constant.

\begin{figure}[htbp] \centering
\includegraphics[angle=0,width=6.0cm]{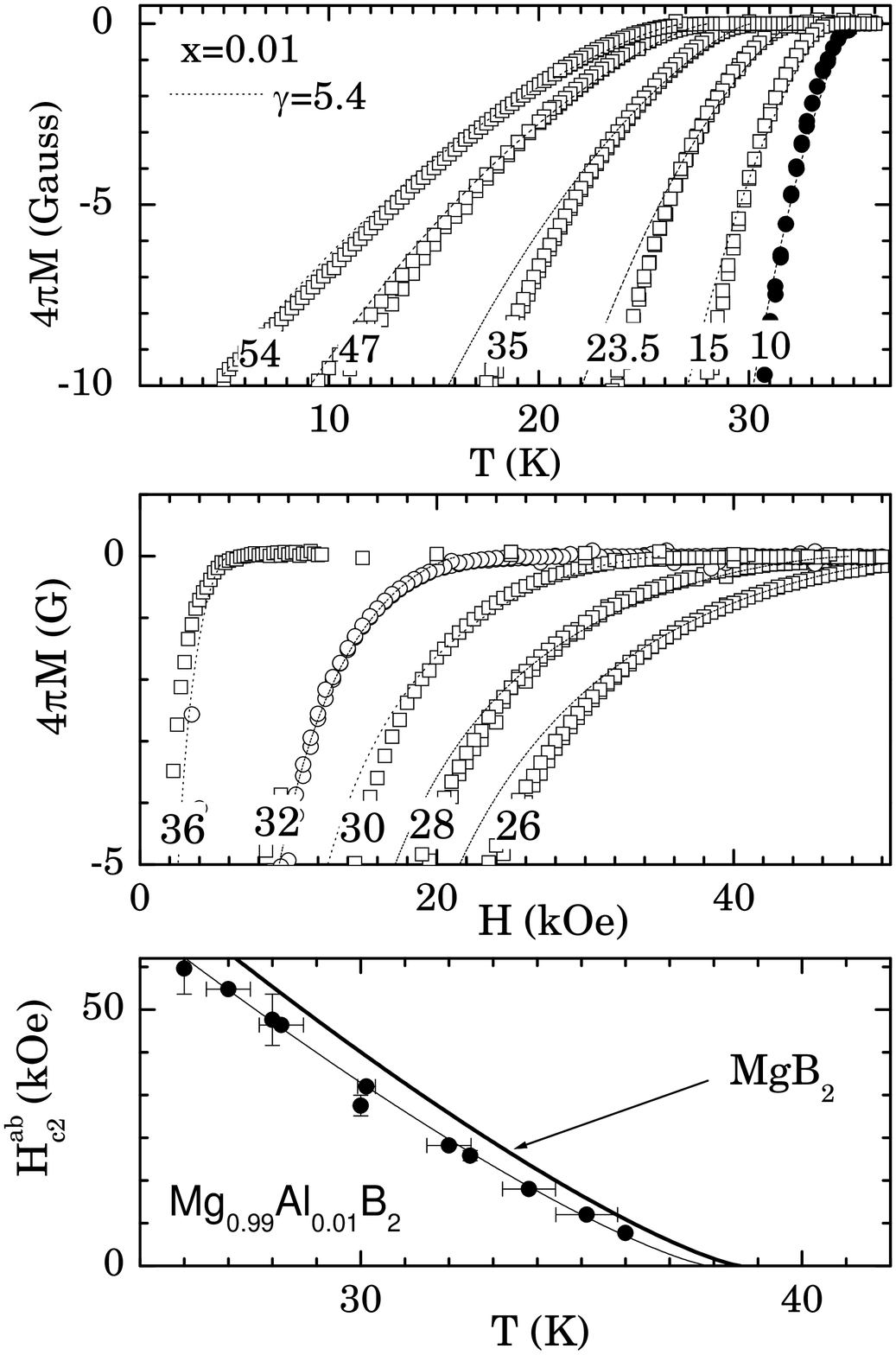}
\caption{
(a) (upper panel) Zero field and field cooling magnetic moment 
as a function of the temperature for $10\leq H\leq 54$ kOe for the powder 
Mg$_{0.99}$Al$_{0.01}$B$_2$ sample used
in the NMR measurements. The dot-lines through the experimental points are simulations
of the reversible magnetic moment supposing an anisotropy $\gamma\approx 5.4$ (see main text).
(b) (middle panel) Isothermal magnetization loops in the reversible regime at 
$26\leq T\leq 36$ K for Mg$_{0.99}$Al$_{0.01}$B$_2$. 
(c) (lower panel) Variation of $H_{c2}^{ab}$ as a function of temperature for the $x=0.01$ sample. 
The solid line is a fit through the experimental points 
with a power law relation $H_{c2}^{ab}=H_{c2}^{ab}(0)(1-T/T_c)^{\nu}$ 
($H_{c2}^{ab}(0)=262\pm 5$ kOe, $T_c=37.9\pm 0.1$ K and $\nu=1.27\pm 0.02$). 
Also included is the $H_{c2}^{ab}(T)$-curve (thick solid line) of the $x=0$ sample, 
for direct comparison.
}
\label{fig3}%
\end{figure}

Dc-magnetic measurements in a SQUID magnetometer under a magnetic field of 
$H=10$ Oe show that all the examined samples are 
superconductors, with their $T_c$ decreasing quasilinearly with increasing Al content  
($dT_c/dx\approx -0.1$ K/at \% Al). A steeper decrease of $T_c$ is observed 
for $x>0.1$ that becomes zero at $x\approx 0.55$.
Figure \ref{fig3} shows the reversible portion of the temperature dependence of the 
magnetization in various fields for $ x=0.01$. 
Contrary to high-temperature superconductors where fluctuation
effects cause a substantial broadening of the transition with increasing temperature and field,
the transition curves for Mg$_{0.99}$Al$_{0.01}$B$_2$ shift to lower temperatures in an almost
parallel fashion. However, instead of the expected conventional linear dependence a pronounced 
curvature is present in $M(T)$ curves. This curvature has been attributed to the 
anisotropy\cite{simon01,budko01b} of MgB$_2$.
The dot-lines through the experimental points is a simulation of the
reversible magnetic moment using the equation\cite{simon01}:
\begin{eqnarray}
4\pi M&=&-\frac{\Phi_{\rm o}}{8\pi\lambda(T)^2\beta_{A}\gamma^{1/3}\sqrt{\gamma^2-1}}\times\nonumber\\
       & &    \left (\frac{1-4h^2}{3h^2}\sqrt{1-h^2}+\ln\left (\frac{1+\sqrt{1-h^2}}{h}\right)\right )\label{eq2}
\end{eqnarray}
where $h=H/H_{c2}^{ab}$, $\lambda=(\lambda_{ab}^2\lambda_c)^{1/3}$ is the average penetration depth, $\beta_{A}=1.16$,
$\Phi_{\rm o}$ is the flux quantum, and $\gamma=H_{c2}^{ab}/H_{c2}^c$ is the anisotropy constant.
In order to simulate the $M(T)$ data we suppose a power law relation for $H_{c2}^{ab}(T)$,
$H_{c2}^{ab}=H_{c2}^{ab}(0)(1-T/T_c)^{\nu}$ 
(with $H_{c2}^{ab}(0)=262\pm 25$ kOe, $T_c=37.9\pm 0.1$ K and $\nu=1.27\pm 0.05$),
a weak magnetic field dependence, $\lambda\sim 2$ nm, and an anisotropy constant $\gamma\sim 5.4$.
Similarly, the $M(H)$ data have been simulated using the same $\gamma$ and a temperature
depended $\lambda(T)$. The value of the anisotropy constant, deduced from magnetic measurements,
for $x=0.01$ is {\it smaller than} that obtained from the $x=0$ sample ($\gamma\sim 6$), 
in agreement with the NMR spectra (vide infra). The temperature variation of $H_{c2}^{ab}$ has 
the same functional form with the $x=0$ sample (i.e. the same exponent), while $T_c$ and 
$H_{c2}^{ab}(0)$ are $0.7$ K and $20$ kOe smaller, respectively.
\begin{figure}[htbp] \centering
\includegraphics[angle=0,width=6.5cm]{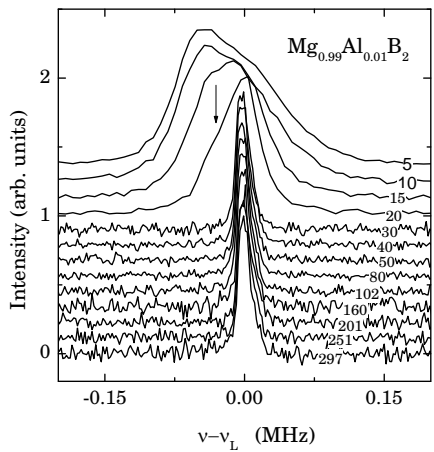}
\caption{$^{11}$B NMR line shapes as a function of temperature
for Mg$_{0.99}$Al$_{0.01}$B$_2$ under a magnetic field $H=23.5$ kOe.
}\label{fig4}%
\end{figure}

$^{11}$B NMR line shape measurements of the central transition ($-1/2\rightarrow 1/2$) 
were performed on two spectrometers operating in external magnetic fields 
$H_{\rm o}=23.5$ and, $47$ kOe. The spectra were obtained 
from the Fourier transform of half of the echo, following a typical 
$\pi/2-\tau-\pi$ spin-echo pulse sequence. 
NMR is a very sensitive local probe of the 
spatially inhomogeneous magnetic field associated with the vortex state, 
which is formed in external magnetic fields $H_{c1}<H_{\rm o}<H_{c2}$ \cite{Fite66,Brandt}. 
In a recent study \cite{Papav} we have obseved that the $^{11}$B NMR line shapes 
in pure MgB$_2$ remain unchanged down to the temperature of the second critical field 
$T_{c2}$ whereas for $T<T_{c2}$, a second peak develops at lower 
frequencies. The intensity ratio of this second peak to 
the unshifted high-T peak was observed to increase in field $H_{\rm o}=23.5$ kOe when compared 
to that in field  $H_{\rm o}=47$ kOe \cite{Papav}. A direct comparison of the NMR line shapes with 
dc-magnetic measurements, that reveal the temperature dependence 
of $H_{c2}^{ab}$ and $H_{c2}^c$, has shown that the intensity and shape of the low frequency 
peak follows the development of the vortex lattice as a function of temperature \cite{Papav} . Since in pure 
MgB$_2$, $H_{c2}^{ab}\approx 150$ kOe \cite{Lima01,simon01}, 
this was explained as showing that a part of the grains remains in the normal state 
(unshifted peak) for $H_{c2}^c<H_{\rm o}<H_{c2}^{ab}$ down to the lowest measured temperature $T=5K$. 
\begin{figure}[tbp] \centering
\includegraphics[angle=0,width=6.0cm]{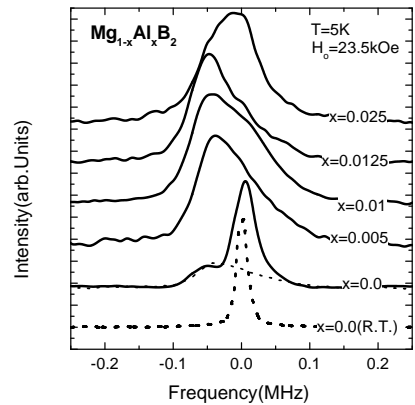}
\caption{
$^{11}$B NMR line shapes of Mg$_{1-x}$Al$_x$B$_2$ 
for $0.0\leq x\leq 0.025$ at $T=5$ K 
in field $H_{\rm o}=23.5$ kOe.}
\label{fig5}
\end{figure}
It is thus remarkable that by light Al doping, the normal state signal component disappears in the mixed 
superconducting state and only the pure vortex lattice signal is present. 
This is clearly seen in Figure \ref{fig4}, which shows the $^{11}$B NMR line shapes 
for $x=0.01$ at various temperatures, in field $23.5$ kOe. Alike to MgB$_2$ spectra 
\cite{Papav}, the line shapes in the normal state are temperature independent. For 
$T\leq 30 $ K the vortex lattice is formed, inducing a gradual shift of the peak frequency 
(corresponding to $H_s$) that creates the characteristic asymmetric 
broadening of the NMR frequency distribution as expected from the 
vortex lattice only.\cite{Brandt} This effect indicates an enhancement 
(relative to pure MgB$_2$) of $H_{c2}^c$ above $23.5$ kOe by Al 
doping, which leaves only the superconducting state at $T=5$ K. At $T=5$ K 
the shift of $H_s$ from the field $H_{\rm o}$ in the normal state is about 
$50$ Gauss. Since we measure an anisotropic polycrystalline sample, it is
expected that the sharp singularities smear out.
It is worth noting that at $T=20$ K the line shape exhibits a shoulder (see arrow in Fig. 
\ref{fig4}). This shoulder indicates that $H_{c2}^{c}$ is crossed at this temperature.

Figure \ref{fig5} shows NMR spectra at $T=5$ K in $H_{\rm o}=23.5$ kOe for 
Mg$_{1-x}$Al$_x$B$_2$ ($0\leq x\leq 0.025$). At $T=300$ K the NMR spectra are essentially 
identical for $0\leq x\leq 0.2$. Since the cell constants change slightly in this 
concentration region, the observed similarity in the NMR spectra indicates that the induced 
line shape is resolution limited. 
At $T=5$ K all the samples are in the mixed state and the line shape
reflects  the magnetic field distribution from the vortex lattice. Remarkably,
the line shapes depend on $x$. As discussed above for the $x=0$ system, the observed line shape
is the result of the anisotropy. Hence, the disappearance of the normal state signal component and the 
variation of the vortex state signal with $x$ can be explained by supposing that the anisotropy decreases 
with Al doping. We stress that even at $x=0.005$ the 
component from the normal state signal disappears. In figure \ref{fig5} we have scaled 
for comparison the signal intensity of the $x=0.005$ system under the low frequency tail 
of pure MgB$_2$. Apparently there is an excellent matching of the two signals, 
providing clear experimental evidence that this shoulder corresponds to the 
magnetic field distribution of the vortex state.
We also notice that for $x\geq 0.025$ the line shape 
changes drastically. At this concentration either the
anisotropy starts to increase abruptly, or the particular line shape 
is associated with the onset of phase separation at this composition.
\begin{figure}[tbp] \centering
\includegraphics[angle=0,width=6cm]{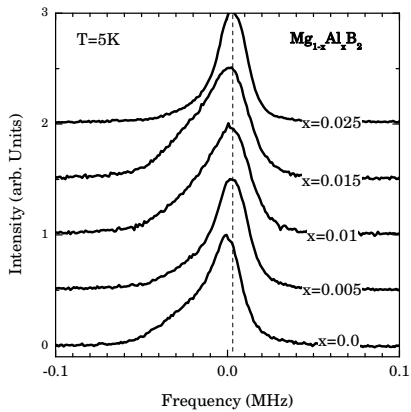}
\caption{$^{11}$B NMR line shapes of Mg$_{1-x}$Al$_x$B$_2$ for 
$0.0\leq x\leq 0.025$ at $T=5$K in a magnetic field 
$H_{\rm o}=47$ kOe.}
\label{fig6}
\end{figure}
Figure \ref{fig6} shows the dependence of the NMR spectra on Al doping for 
Mg$_{1-x}$Al$_x$B$_2$ ($0\leq x\leq 0.025$), at $T=5$ K and $H_{\rm o}=47$ 
kOe. Contrary to Figure \ref{fig5}, in all samples the line shapes 
exhibit a low frequency tail and an unshifted peak, corresponding to coexisting vortex 
and normal-state components. This indicates that $3.2\leq \gamma \leq 6.4$ (e.g for $x=0.01$ sample) 
by considering $H_{c2}^{ab}\approx 150$ kOe. 

The observed decrease of anisotropy can be attributed to the progressive electron 
filling of the $\sigma$ bands with increasing $x$, which reduces the anisotropy of the boron $p$ states. In the 
microscopic theory\cite{gorkov63} the anisotropy parameter is given by 
$\gamma^2=\langle \Delta({\bf k}_F) v_{ab}^2\rangle/\langle\Delta({\bf k}_F) v_{c}^2\rangle$, 
where $v_i$ are the Fermi velocities and $\langle\cdots \rangle$ stands for Fermi surface 
averages. When the ratio $\langle v_{ab}^2\rangle/\langle v_{c}^2\rangle$ is averaged over 
the entire Fermi surface for MgB$_2$ it is close to unity\cite{kortus01,budko01b}, which means 
a strong anisotropy of $\Delta({\bf k}_F)$. Following the arguments of Bud'ko {\it et 
al.}\cite{budko01b}, the electron-phonon interaction is particularly strong on 
the Fermi surface sheets which are shaped as slightly distorted cylinders along the 
$c$ crystal direction. If the gap $\Delta$ on the remaining
Fermi surface sheets is negligible, the reduction of the anisotropy could originate 
from the reduction of the $\sigma$-holes, as we mentioned above.

In conclusion, we show for the first time the magnetic field distribution in the pure vortex state of lightly doped 
Mg$_{1-x}$Al$_x$B$_2$ by using $^{11}$B NMR line shape measurements. Our NMR and magnetization data 
reveal that substitution of Al for Mg reduces the anisotropy substantially, apparently due to reduction of 
the $\sigma$-holes. This shows up the important role of $p_{x,y}$ orbitals (which form  the 
$2D$ $\sigma$-holes band) in the superconductivity of MgB$_2$. We argue that our results provide an 
experimental basis for further theoretical investigations concerning the 
role of the $\sigma$-bands in the superconducting mechanism of MgB$_2$.

We thank the ESRF for provision of synchrotron X-ray beamtime and P. Pattison 
and I.A. Beukes for help with the experiments.


\begin{references}
\bibitem{Comment1} Images of the Debye-Scherrer rings were recorded on the $345$ mm diameter
Mar Research circular image plate system on the BM1A beamline at the 
ESRF,
Grenoble. A monochromatic X-ray beam of 
wavelength, $\lambda$= $0.79983$ \AA\ and dimensions $0.5\times0.5$ mm$^2$ was focussed
onto the sample by sagittal bending of the second crystal of the double-crystal Si
(111) monochromator. Patterns were measured with sample-to detector
distance of $200$ mm for periods of $10$ s. During the data
collection the sample was rotated about its axis by $10^{\rm o}$. 
One-dimensional diffraction patterns were obtained by integrating around
the rings using local software (program FIT2D).
\end{references}
\end{document}